# Mircomechanical insights into unconstrained grain boundary sliding

Divya Sri Bandla[1], Subin Lee[1,*], Christoph Kirchlechner[1]

[1] Institute for Applied Materials, Karlsruhe Institute of Technology, Karlsruhe 76131, Germany

* Corresponding author: Subin Lee (subin.lee@kit.edu)

## Abstract

Grain boundary sliding is a key deformation mechanism at high homologous temperatures in polycrystalline materials, however, its intrinsic behavior is often obscured by strain accommodation processes. In this study, dislocation-mediated unconstrained GBS was investigated using Ni bicrystal micropillars containing a high-angle grain boundary. Micropillar compression tests were performed from room temperature to 600 °C and strain rates between $5\times10^{-4}$ and $10^{-1}$ $s^{-1}$. By comparing bicrystal and single-crystal responses, intrinsic contribution of grain boundary sliding was isolated. The strain-rate sensitivity remained low (0.034 ± 0.017), comparable to room temperature values, indicating the absence of diffusion-controlled accommodation mechanisms. The activation energy for grain boundary sliding was determined to be 284 ± 47 kJ $mol^{-1}$, consistent with lattice diffusion-assisted dissociation of lattice dislocations into grain boundary dislocations. These results demonstrate that the high strain-rate sensitivity associated with grain boundary sliding in polycrystals originates primarily from accommodation processes rather than the intrinsic sliding mechanism.

Grain boundary sliding (GBS) is a phenomenon in polycrystalline materials, where the adjacent grains slide relative to each other under an applied stress [1,2]. It is typically characterized by a high strain-rate sensitivity (*SRS* ≈ 0.3–0.5), and an activation energy

*Q* comparable to that of lattice or grain boundary diffusion [3,4]. However, GBS in polycrystalline materials causes compatibility stress at grain boundary triple junctions and therefore requires an accommodation mechanism, such as dislocation climb, to relieve stress concentrations and further progress of GBS [5]. Consequently, when deformation is limited by accommodation mechanisms, the experimentally measured *SRS* and *Q* values reflect the accommodation processes rather than intrinsic GBS, rendering a thorough understanding of GBS in polycrystalline materials challenging.

A quantitative understanding of intrinsic GBS requires deformation parameters under unconstrained conditions, i.e., in the absence of accommodation mechanisms. This is particularly relevant for grain boundary engineering approaches, such as solute segregation or boundary character control, where isolating the intrinsic GBS characteristics is essential. In addition, the diversity of grain boundary types in polycrystals complicates systematic investigation of the effect of boundary character on GBS behavior.

Small-scale mechanical testing provides an effective approach to overcome these limitations by enabling the isolation of specific microstructural features. Bicrystal micropillars containing a single grain boundary can be fabricated using focused ion beam (FIB) milling, eliminating constraints from triple junctions and allowing direct investigation of unconstrained GBS. Previous studies on β-Sn [6] and Al [7] demonstrated GBS in such micropillars at room temperature (RT) due to the low melting points, quantifying strain contributions of GBS, as well as size effects. However, deformation parameters governing unconstrained GBS have not yet been fully established.

In our recent work, we demonstrated unconstrained GBS in Ni bicrystal micropillars containing a random high-angle grain boundary (HAGB) at 250–300 °C [8]. Based on the observations of dislocation activities along the GB and inverse dependence of activation stress on pillar diameter, the sliding mechanism was identified as dislocation-mediated.

In this study, micropillar compression tests were conducted over a temperature range from RT to 600 °C to quantify the *SRS* and *Q* of dislocation-mediated unconstrained GBS. Complementary experiments were performed on single-crystal micropillars under identical conditions. The comparison of the deformation response and the corresponding

*SRS* and *Q* values between bicrystal and single-crystal pillars enables isolation of the intrinsic GBS contribution and provides a quantitative basis for assessing the underlying rate-controlling mechanisms.

A Ni bicrystal containing a random HAGB of misorientation 20° was grown by the Bridgman method. A small piece measuring 7×3×2 mm was mechanically polished and vibropolished prior to micropillar fabrication. The normal directions of the two grains were $[344]$ and $[324]$, and the grain boundary was inclined by approximately 35° relative to the sample surface (Fig. 1a). Micropillars with a diameter of 1 and 3 μm (aspect ratio 2–2.5) were fabricated using Ga-FIB operated at 30 keV (Crossbeam 550L, Zeiss). The detailed milling parameters are provided in [8]. *In situ* SEM pillar compression tests were performed using a nanoindenter (Hysitron PI89, Bruker) in a displacement-controlled mode. Strain rates ranged from $5\times10^{-4}$ to $10^{-1}$ $s^{-1}$, and the temperature from RT to 600 °C. At each test condition, 7–10 pillars were tested. The loading direction was parallel to the pillar axis, as shown by the black arrow in Fig. 1a. Flat punches with diameters of 2 and 5 μm (Synton-MDP) were used to test the 1 and 3 μm pillars, respectively. Diamond punches were used for RT testing, while tungsten carbide punches were used for elevated temperature testing. The target temperature was achieved at a heating rate of 10 °C $min^{-1}$, and the temperature remained stable during the test within ± 0.1 °C.

Our previous work identified a critical temperature for sliding along the present HAGB between 250 and 300 °C [8]. Accordingly, 300 °C was selected to evaluate the *SRS* under GBS conditions, while RT was used as a reference without GBS. Figure 1 illustrates the activation of GBS at ~300 °C. Representative deformed Ni bicrystal micropillars ($d$ = 1 μm) tested at RT and 300 °C at a strain rate of $10^{-2}$ $s^{-1}$ are shown in Figs. 1b and 1c. Slip traces are observed in both grains, as indicated by colored dashed lines. In the [324] grain, the activated slip systems were identified as $(1\bar{1}1)[011]$ and $(\bar{1}11)[101]$ with Schmid factor being 0.39 and 0.36, respectively. In the [344] grain, slip occurred along $(1\bar{1}1)[011]$ and $(\bar{1}11)[110]$ with Schmid factors of 0.31 and 0.25, respectively. However, at 300 °C, in addition to slip traces, clear evidence of GBS is observed as indicated by a white dashed ellipse in Fig. 1c, where the grain [324] was sheared with respect to the [344] grain along the HAGB. Similarly, GBS was also observed at 300 °C and other strain rates of 5×$10^{-4}$, $10^{-3}$, and $10^{-1}$ $s^{-1}$ (see supplementary material).

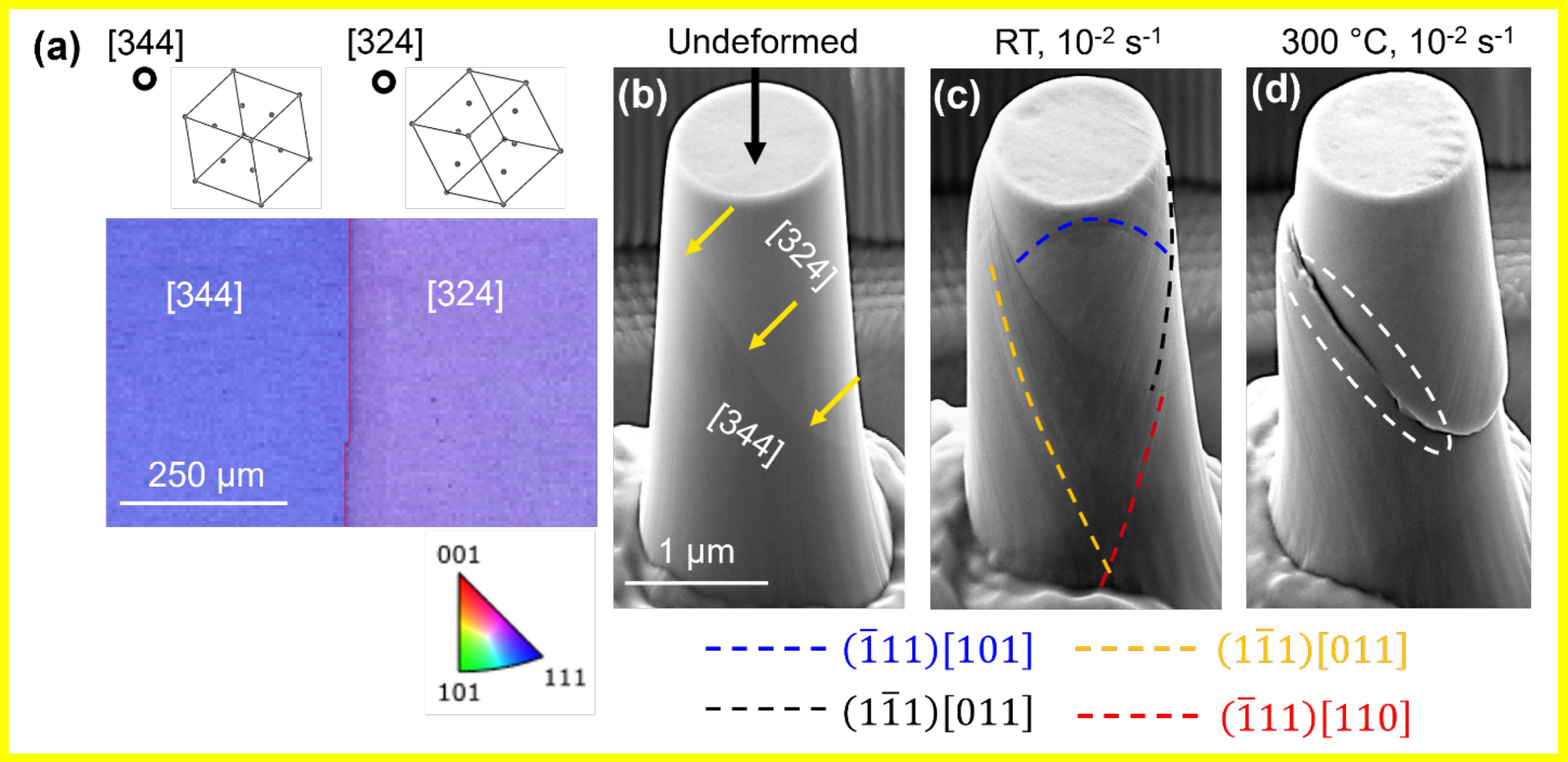

**Fig. 1.** (a) EBSD inverse pole figure map (in Z-direction) of bicrystal sample and unit cells of two grains. Secondary electron images of (b) a fabricated bicrystal micropillar (*d* = 1 μm), highlighting the HAGB with yellow arrows and the micropillar compression loading direction with a black arrow, and (c) and (d) are pillars deformed to a strain of 0.1 at RT and 300 °C, with a strain rate of $10^{-2}$ $s^{-1}$. The colored dashed lines in (c) indicate the slip traces, and the white dashed ellipse in (d) indicates the activity of GBS.

Representative stress-strain curves obtained at RT and 300 °C for different strain rates are shown in Figs. 2a and 2b, respectively. An intentional unloading segment at 5% strain was introduced to minimize effects of lateral constraints during compression. The yield stress at 300 °C is lower than at RT, while in both cases, the yield stress at 5% plastic strain increases with the strain rate, indicating a positive *SRS*. The *SRS* was quantified from the slope of the 5% yield stress versus strain rate data on a double-logarithmic scale. The slope of the linear fit to the data represents the *SRS* (see Fig. 2c). The resulting *SRS* values are 0.047 ± 0.020 and 0.034 ± 0.017 at RT and 300 °C, respectively; the reported error bars correspond to the standard error of the fitted slope.

Since the *SRS* is sensitive to the grain size [9,10], our results for 1 μm bicrystal pillars are compared with polycrystalline Ni with a 1 μm grain size. The *SRS* at RT in polycrystals (~0.007 [9]) is significantly lower than in the present study. A similar trend has been reported for Cu, where micropillars exhibited a higher *SRS* than bulk samples due to differences in rate-controlling mechanism [11,12]. In bulk polycrystals, deformation is governed by dislocation glide and its interaction with obstacles, such as grain boundaries and other dislocations, whereas, in micropillars, dislocation multiplication controls the deformation due to the limited sample volume [13]. The *SRS* measured here is consistent with reported values for Ni single-crystal micropillars at RT, which is approximately 0.037 [14].

At the elevated temperature (300 °C) where GBS is active, the *SRS* remains comparable to that at RT (see Fig. 2c). This contradicts the classical observation that *SRS* values are high during GBS, for instance, ~0.8 for electrodeposited Ni with a grain size of ~0.68 μm [3] and ~0.5 for polycrystalline Al with a grain size of ~10–20 μm [15]. In polycrystals, dislocation-mediated GBS sliding requires accommodation by dislocation climb [5], making the overall deformation rate sensitive to vacancy diffusion and thus increasing *SRS*. In contrast, unconstrained GBS in bicrystal micropillars does not require the activation of additional accommodation mechanisms. As a result, deformation remains governed by dislocation processes similar to those at RT, leading to comparable *SRS* values. This interpretation is consistent with observations by Feldner *et al.* [16], who reported a breakdown in high *SRS* for 1 μm pillars of Zn-22% Al.

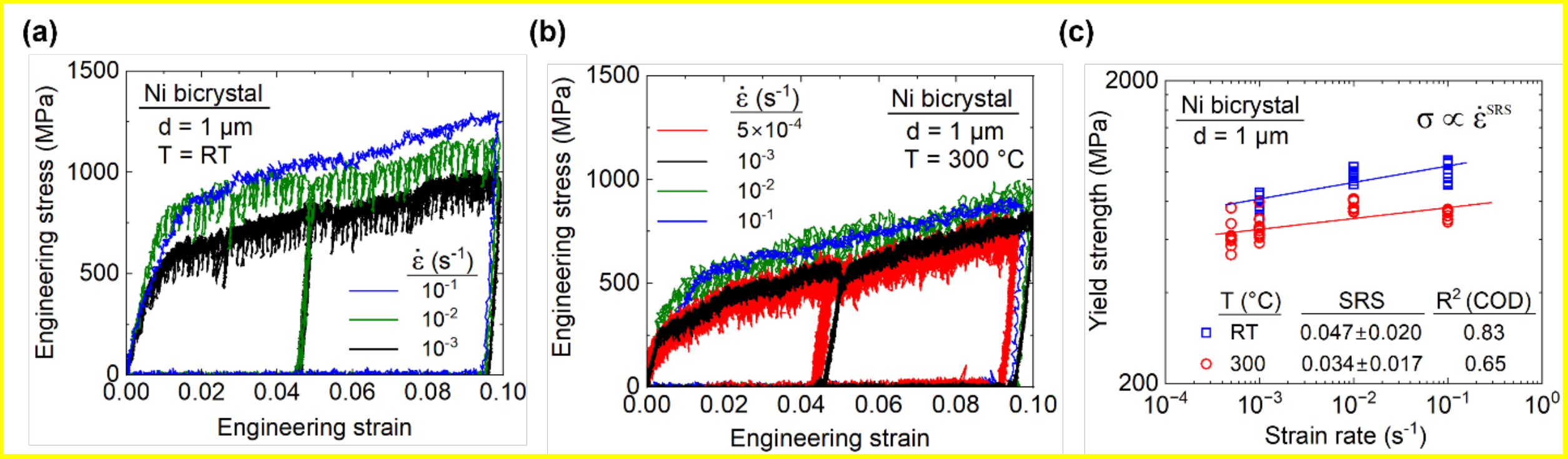

**Fig. 2.** Representative stress-strain curves of 1 μm pillars deformed at (a) RT and (b) 300 °C with strain rates from $10^{-1}$ to $5 \times 10^{-4}$ $s^{-1}$. (c) is the determination of *SRS* at RT and 300 °C, where each data point is from an individual micropillar compression test. The fit quality parameter ($R^2$), coefficient of determination (COD), is also specified.

The activation energy, *Q*, of unconstrained GBS, was determined from bicrystal micropillar compression over a range of temperatures at a strain rate of $10^{-3}$ $s^{-1}$. GBS was observed in the temperature range of 250 to 600 °C. Representative micrographs of deformed pillars within the GBS regime are shown in Figs. 3a to 3d. At lower temperatures within this range (up to 400 °C), slip traces are visible in both grains, corresponding to $(\bar{1}11)[101]$ in [324] grain and $(1\bar{1}1)[011]$ in [344], as highlighted by colored dashed lines in Fig. 3a and Fig. 3b. At higher temperatures (500 and 600 °C, shown in Fig. 3c and 3d, respectively), however, the slip traces are no longer visible, possibly due to oxide scale formation on the pillar surface.

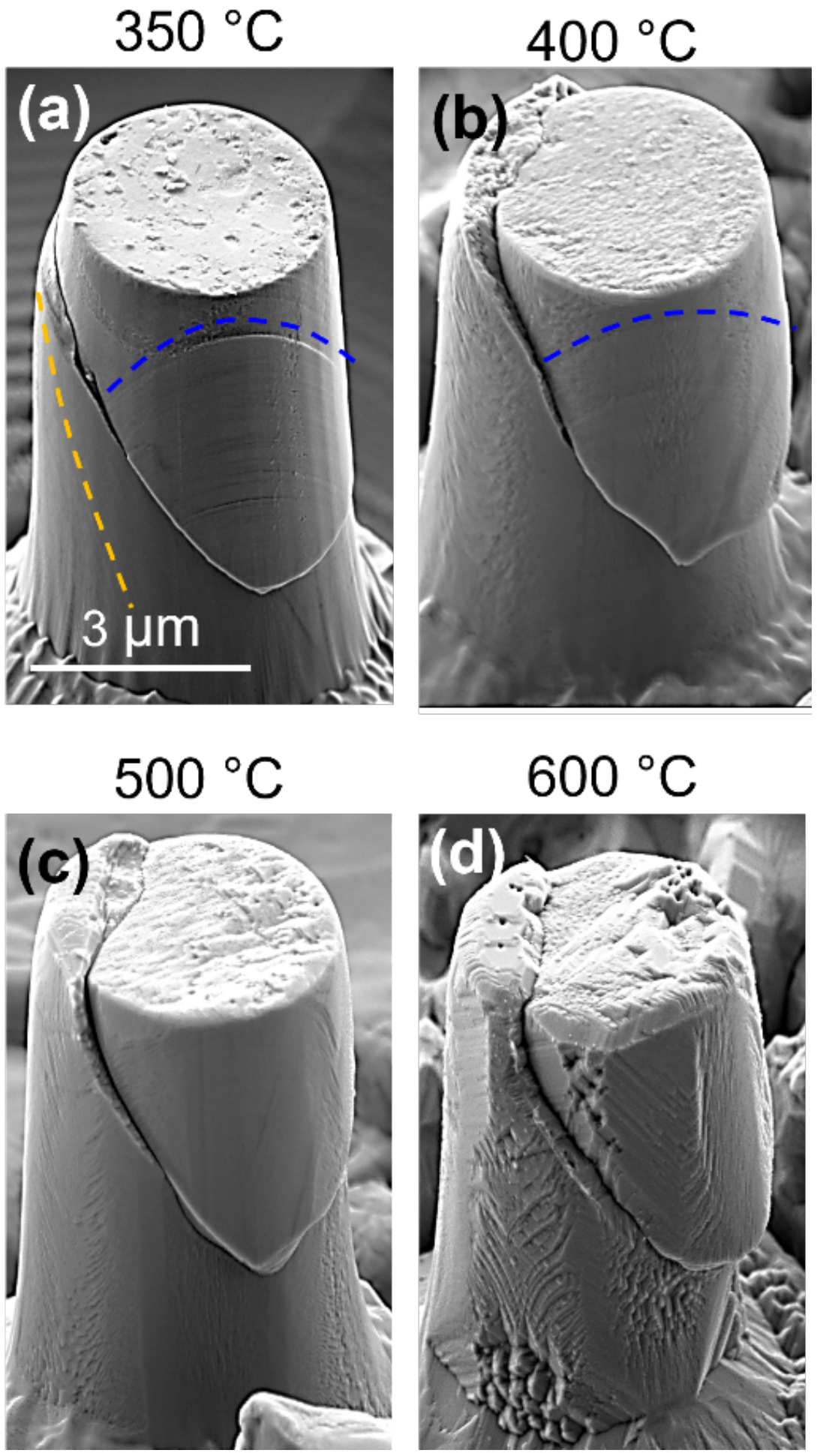

**Fig. 3.** Secondary electron images of 3 μm pillars deformed to 0.1 strain with $10^{-3}$ $s^{-1}$ strain rate at (a) 350, (b) 400, (c) 500, and (d) 600 °C. The slip traces in (a) are highlighted with colored dashed lines.

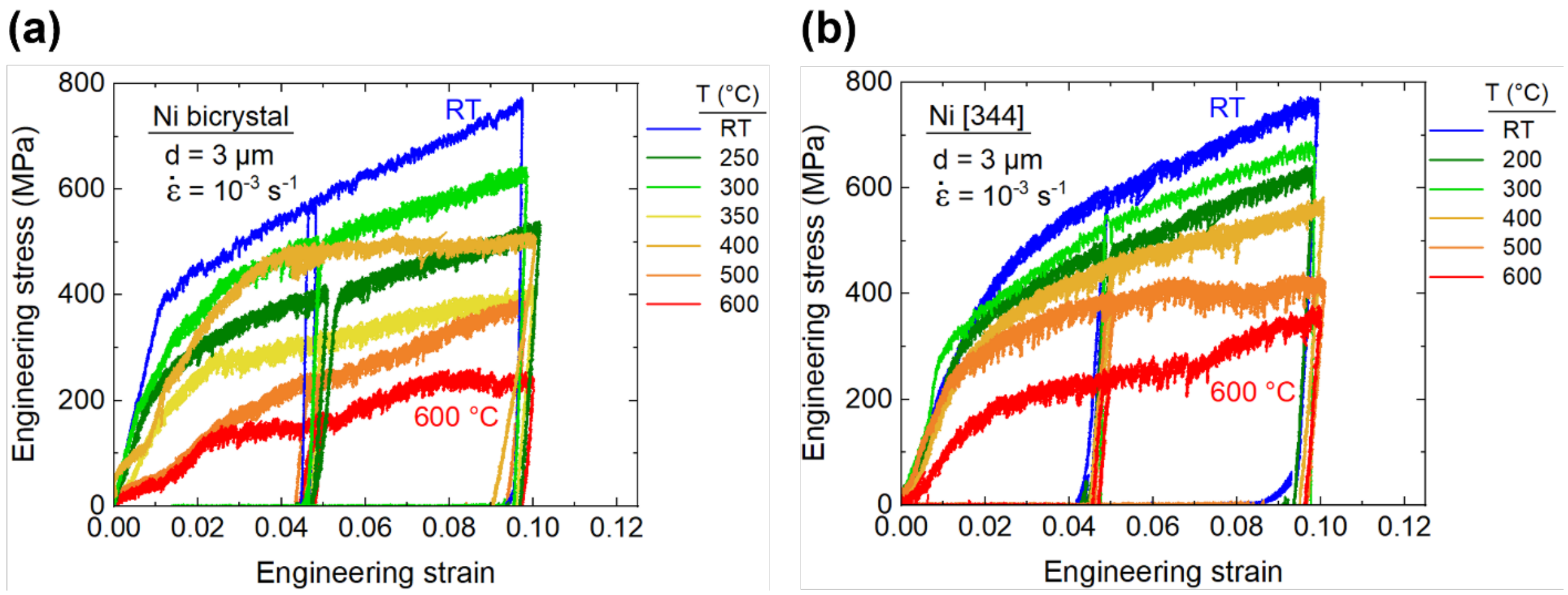

**Fig. 4.** Engineering stress-strain curves of 3 μm (a) bicrystal and (b) [344] pillars deformed at temperatures from RT to 600 °C with a strain rate of $10^{-3}$ $s^{-1}$.

Fig. 4a shows the typical stress-strain curves for bicrystal micropillars (*d* = 3 μm) deformed from RT to 600 °C at a strain rate of $10^{-3}$ $s^{-1}$. The yield stress decreases with increasing temperature despite minor fluctuations. For comparison, identical tests were carried out on the single crystalline [344] micropillars (Fig. 4b), which exhibit a similar temperature dependence of yield stress – i.e. a reduction of the yield stress with increasing temperature. The yield stress values from the stress-strain curves of [344] single-crystal and bicrystal micropillars are plotted against the inverse of temperature to determine *Q* (Fig. 5). According to the Arrhenius-type constitutive equation, at a constant strain rate, the flow stress and temperature are related as [17],

$$\dot{\varepsilon} \propto \sigma^n exp\left(\frac{-Q}{RT}\right) \quad (1)$$

where *R* is a gas constant (8.314 J $mol^{-1}$ $K^{-1}$), and *n* is the stress exponent, which can also be considered as the inverse of *SRS*. After rearranging the equation,

$$\sigma^{1/SRS} \propto exp\left(\frac{Q}{RT}\right) \quad (2)$$

Therefore, on a semi-logarithmic plot of yield stress versus the inverse of temperature, the slope of the data represents the activation energy. Owing to the oxidation of the micropillar surface at 600 °C (Fig. S2), the corresponding data were excluded from the *Q* calculations for both single-crystal and bicrystal pillars.

Following the literature data on single-crystal pillars of Ni [14], a constant *SRS* value of 0.037 was assumed over the temperature range from RT to 500 °C for [344] pillars. For bicrystal pillars, *SRS* determined in the non-GBS ($SRS_{RT}$ = 0.047) and GBS ($SRS_{300\ °C}$ = 0.034) regions were used for the RT–250 °C and 250 °C–500 °C temperature ranges, respectively.

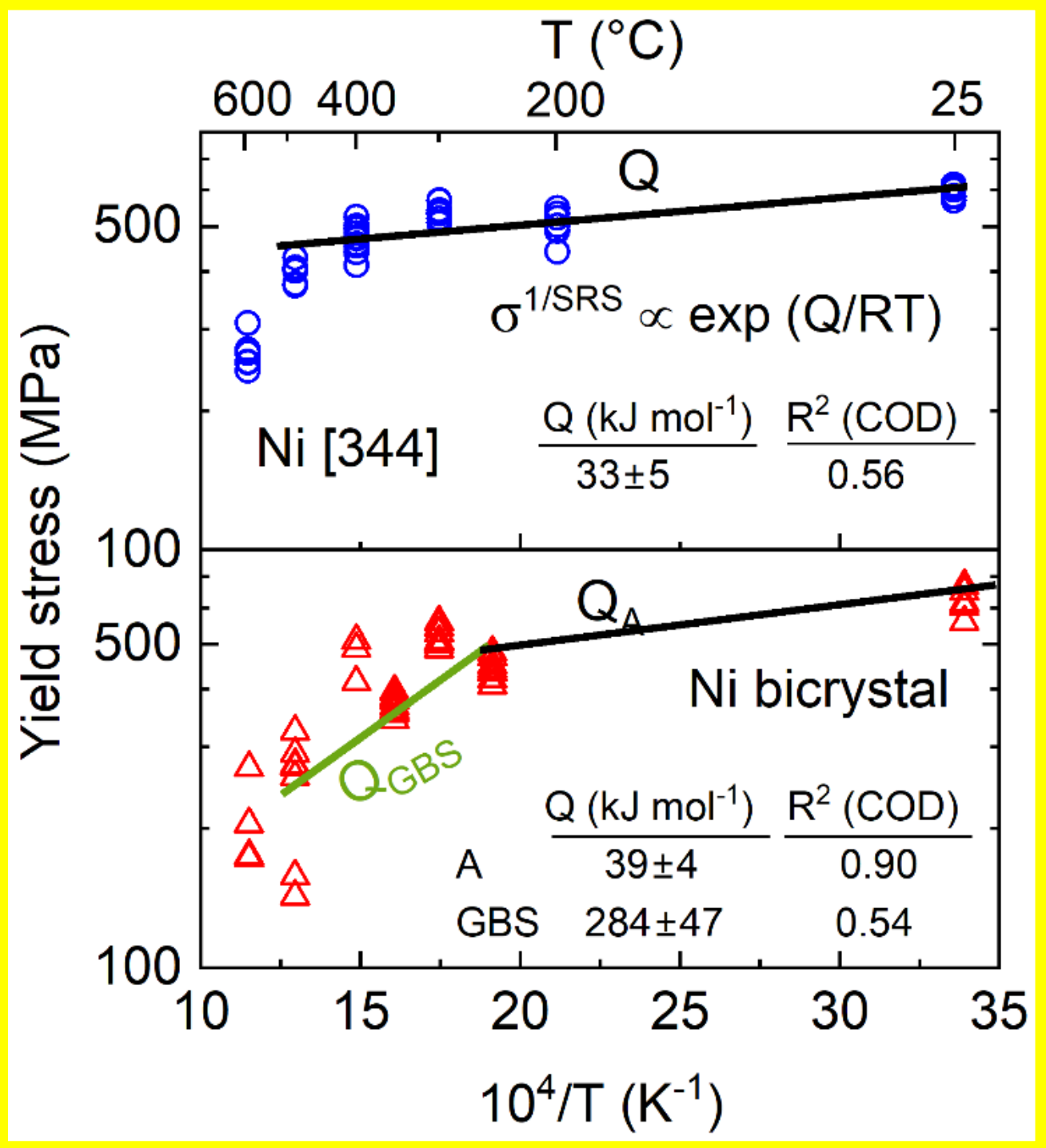

**Fig. 5.** Activation energy determination of Ni [344] single-crystal and bicrystal micropillars of diameter 3 μm. Each data point is from an individual micropillar compression test.

As shown in Fig. 5, [344] single-crystal pillars exhibit a gradual decrease in yield stress as temperature increases from RT to 500 °C, yielding an activation energy of 33 ± 5 kJ mol$^{-1}$. In contrast, the bicrystal micropillars do not exhibit a single deformation mechanism over the entire temperature range. As shown in Fig. 1 and Fig. 3, their deformation mechanism transitions from in-grain slip to GBS at 250 °C, therefore, the activation energy of these two mechanisms is calculated separately at two temperature regimes. Accordingly, the bicrystal pillars yield activation energies of 39 ± 4 and 284 ± 47 kJ mol$^{-1}$ in the temperature ranges of RT–250 °C and 250–500 °C, respectively.

At low temperatures (RT to 500 °C for single-crystals and up to 250 °C for bicrystals until GBS activates), the measured activation energies are consistent with dislocation glide as the rate-controlling mechanism [18]. The slightly higher value observed in micropillars compared to bulk FCC metals can be attributed to size effects, particularly the influence of free surfaces and limited dislocation sources. In bicrystal, the marginally higher activation energy observed compared to the single-crystal pillars, likely arises from the presence of a grain boundary, which acts as an additional barrier to the dislocation motion.

The behavior at elevated temperatures, where GBS is active, is of primary interest. The activation energy obtained for bicrystal micropillars (284 ± 47 kJ mol$^{-1}$) is similar to that of the lattice diffusion process in Ni ($Q_L$ = 278 kJ mol$^{-1}$) [19]. We already observed in our recent work that the GBS along the current HAGB in Ni is a dislocation-mediated mechanism, [8]. Essentially, the dislocation-mediated GBS involves sequential processes of glide of lattice dislocations to the grain boundary, dissociation of lattice dislocations into grain boundary dislocations, and glide of grain boundary dislocations along the grain boundary. This suggests that one of these processes involves lattice diffusion-assisted processes, thus resulting in the $Q$ of 284 ± 47 kJ mol$^{-1}$. We speculate that the dissociation of lattice dislocations into grain boundary dislocations may be limited by a lattice diffusion dominated process.

In summary, unconstrained grain boundary sliding was investigated in Ni bicrystal micropillars over a range of temperatures and strain rates. The measured strain-rate sensitivity remained low ($SRS$ ≈ 0.034 ± 0.017), indicating that, in the absence of accommodation constraints, GBS is governed by dislocation-mediated processes rather than diffusion-controlled mechanisms. The activation energy $Q$ for GBS was determined to be 284 ± 47 kJ mol$^{-1}$, which is attributed to lattice diffusion-controlled assisted dissociation of lattice dislocations into grain boundary dislocations at the grain boundary. Overall, the results demonstrate that the high strain-rate sensitivity commonly observed in polycrystalline GBS originates primarily from accommodation processes, whereas the intrinsic GBS mechanism is controlled by dislocation activity at grain boundaries.

## Acknowledgment

The authors express their gratitude to the Deutsche Forschungsgemeinschaft (DFG) for its financial support within project 500076185 “Micromechanical characterization of grain boundary slip: Towards a deformation mechanism map”.

## Declaration of competing interest

No potential conflict of interest was reported by the author(s).

## Data availability

Datasets supporting this study will be deposited on Zenodo upon publication of the article.